# Deviance Detection and Regularity Sensitivity in Dissociated Neuronal Cultures

Zhuo Zhang[1*], Amit Yaron[2*], Dai Akita[1], Tomoyo Isoguchi Shiramatsu[1], Zenas C. Chao[2], Hirokazu Takahashi[1,2**]

[1]Department of Mechano-Informatics, Graduate School of Information Science and Technology, The University of Tokyo, Tokyo, Japan

[2]International Research Center for Neurointelligence (WPI-IRCN), UTIAS, The University of Tokyo, Tokyo, Japan

*These authors have contributed equally to this work and share the first authorship

** Correspondence:
Hirokazu Takahashi
takahashi@i.t.u-tokyo.ac.jp

## Abstract

Understanding how neural networks process complex patterns of information is crucial for advancing both neuroscience and artificial intelligence. To investigate fundamental principles of neural computation, we studied dissociated neuronal cultures, one of the most primitive living neural networks, on high-resolution CMOS microelectrode arrays and tested whether the dissociated culture exhibits regularity sensitivity beyond mere stimulus-specific adaptation and deviance detection. In oddball electrical stimulation paradigms, we confirmed that the neuronal culture produced mismatch responses (MMRs) with true deviance detection beyond mere adaptation. These MMRs were dependent on the N-methyl-D-aspartate (NMDA) receptors, similar to mismatch negativity (MMN) in humans, which is known to have true deviance detection properties. Crucially, we also showed sensitivity to the statistical regularity of stimuli, a phenomenon previously observed only in intact brains: the MMRs in a predictable, periodic sequence were smaller than those in a commonly used sequence in which the appearance of the deviant stimulus was random and unpredictable. These results challenge the traditional view that a hierarchically structured neural network is required to process complex temporal patterns, suggesting instead that deviant detection and regularity sensitivity are inherent properties arising from the primitive neural network. They also suggest new directions for the development of neuro-inspired artificial intelligence systems, emphasizing the importance of incorporating adaptive mechanisms and temporal dynamics in the design of neural networks.

## 1 Introduction

Understanding how neural networks process and detect complex patterns of information is fundamental to both neuroscience and artificial intelligence. The brain has a remarkable ability to discriminate between relevant and irrelevant stimuli. In complex environments, the brain responds weakly to repeated unimportant stimuli or background noise, while responding strongly to important stimuli such as sudden alarms. A growing body of evidence suggests that the excellent sensitivity to incoming stimuli that violate learned expectations results not only from passive stimulus-specific adaptation (SSA) but also from an active deviant detection system (Harpaz et al., 2021; Malmierca & Auksztulewicz, 2021; Taaseh et al., 2011).

The evolutionary conservation of deviance detection across species highlights its fundamental importance for survival. Deviance detection has been extensively studied across multiple sensory modalities in humans (Grimm et al., 2011, 2016; Ishishita et al., 2019; Recasens et al., 2014) and in various animal models, including cats (Ulanovsky et al., 2003), rodents (Shiramatsu et al., 2013; Harms et al., 2014; Chen et al., 2015; Pérez-González et al., 2021), bats (Wetekam et al., 2024), and non-human primates (Fishman & Steinschneider, 2012; Takaura & Fujii, 2016). These studies demonstrated various forms of regularity detection beyond deviance detection in simple oddball paradigms, such as in local-global patterns (Strauss et al., 2015; Wacongne et al., 2012), complex statistical regularities (Dürschmid et al., 2016; Yaron et al., 2012), abstract rule violations (Bendixen et al., 2009, 2012), and hierarchical sequence structures (Barascud et al., 2016; Southwell & Chait, 2018). In the human brain, mismatch negativity (MMN), an event-related potential (ERP) in the 150–250 ms post-stimulus latency, has been most extensively characterized as a neural signature of deviant detection (Näätänen et al., 1978; Ross et al., 2020; Fitzgerald et al., 2020; Näätänen et al., 2019). In rats, similar mismatch responses (MMR) have typically been observed between 29-125 ms post-stimulus latency (Ahmed et al., 2011; Nakamura et al., 2011; Astikainen et al., 2011; Shiramatsu et al., 2013; 2020; Shiramatsu and Takahashi, 2021)), considerably earlier than MMN in humans (Näätänen et al., 2012). Therefore, the time scale for deviant detection is likely to differ between species.

The N-methyl-D-aspartate (NMDA) receptor system plays a critical role in deviant detection and MMN generation through its involvement in synaptic plasticity and learning. Extensive evidence links NMDA receptors to MMN generation, as receptor blockade has consistently attenuated MMN/MMR responses across species (Javitt et al., 1996; Rosburg et al., 2016; Shiramatsu et al. 2013; Shiramatsu and Takahashi, 2021). In addition, the involvement of NMDA receptors is relevant to the broader neuronal mechanisms underlying deviance detection. Specific doses of NMDA receptor antagonists have been demonstrated to impair deviance detection and reduce the latency range of this process. This effect has been observed across diverse models, ranging from humans (Tikhonravov et al., 2010; de la Salle et al., 2019) to rodents (Ehrlichman et al., 2008; Harms et al., 2018), suggesting a conserved molecular mechanism underlying deviance detection. NMDA receptor dysfunction appears to primarily affect the efficiency of synaptic transmission by reducing the burst activity that normally enables effective communication between cortical neurons (Jackson et al., 2004).

Considering that deviant detection is one of the most primitive neural processing mechanisms for complex pattern inputs, the dissociated culture of neurons provides an ideal experimental platform for investigating deviance detection mechanisms. Isolated neurons plated on a microelectrode array develop a neural network in a self-organized manner (Potter and DeMarse, 2001; Marom and Shahaf, 2002; Yada et al., 2016; 2017) with remarkable capabilities including pattern recognition (Dranias et al., 2013), adaptive learning (Shahaf and Marom, 2001), goal-directed behavior in embodied systems (DeMarse et al., 2001; Bakkum et al., 2008; Tessadori et al., 2012; Kagan et al., 2022; Yada et al., 2021; Masumori et al., 2020), and predictive coding computations aligned with free energy minimization (Isomura et al., 2015, 2023; Isomura and Friston, 2018; Lamberti et al., 2023, 2024). Early evidence for differential processing of frequent and infrequent stimuli came from Eytan et al. (2003), who demonstrated selective adaptation in these networks, although their data likely primarily reflected SSA. More recently, Kubota et al. (2021) provided initial evidence for true deviance detection using high-density CMOS microelectrode arrays with oddball and many-standards control (MSC) paradigms.

Substantially extending these initial observations, we hypothesized that dissociated neuronal cultures exhibit regularity sensitivity beyond mere SSA and deviance detection. We cultured cortical neurons from the dissected cortex of rat embryos on CMOS arrays, which allowed high-resolution temporal and spatial measurements of spontaneously growing neurons in response to controlled electrical stimulation patterns. Using oddball paradigms across multiple cultures and experimental conditions, we first confirmed MMRs, particularly in the late phase (11-150 ms post-stimulus latency), with true deviance detection beyond mere adaptation. Similar to MMRs in vivo, we showed that these MMRs are NMDA receptor-dependent. We then demonstrated that cultured networks can encode complex temporal patterns and discriminate between random and periodic sequences. This suggests that sensitivity to statistical regularities is a fundamental organizing principle of neural circuits that emerge spontaneously during network development, rather than a sophisticated computational capability that requires intact brain circuits, as previously thought.

## 2 Materials and methods

### 2.1 Preparation of Neuronal Culture

All experimental procedures followed the ethical standards of the Japanese Physiological Society and the protocol was approved by the Ethics Committee of the Graduate School of Information Science and Technology, the University of Tokyo (JA21-8). Cortical neurons were harvested from embryonic day 18 Wistar rats that were euthanized under deep anesthesia to ensure humane treatment. A total of six cerebral cortices were dissected from the left and right brains of three rat embryos per dissection (see Figure 1A for the locations of the cerebral cortices used). The cerebral cortices were enzymatically dissociated in 0.25% trypsin-EDTA (Thermo Fisher Scientific) and then mechanically dissociated in the culture medium using a pipette.

The density of the neuronal cells was adjusted to 38,000 cells per 5 μL. This 5-μL cell suspension was applied directly to the surface of a high-density CMOS microelectrode array (MaxOne, MaxWell Biosystems), which had been pretreated with 0.05% polyethyleneimine (PEI, Sigma-Aldrich) and laminin (Sigma-Aldrich) to enhance cell adhesion to the electrode surface. After plating, 0.6 mL of Neurobasal medium was added to adequately cover the cells. Cultured cells were stored in an incubator at an internal temperature of 36.5°C and a carbon dioxide concentration of 5% to maintain optimal environmental conditions for cell growth.

To maintain cell viability, half of the medium was replaced the next day with a growth medium consisting of 450 mL of DMEM (Thermo Fisher Scientific), 50 mL of horse serum (Cytiva), 1.25 mL of 0.5 mM GlutaMAX (Thermo Fisher Scientific), and 5 mL of 1 mM sodium pyruvate (Thermo Fisher Scientific) (Bakkum et al., 2013; Yada et al., 2016, 2017; Kubota et al., 2019; Ikeda et al.,2025), followed by medium changes were performed every 3-4 days. Cultures were maintained in vitro for 31-35 days to allow for neuronal maturation and stable synaptic connectivity before experimental procedures were initiated.

### 2.2 Electrical Stimulation and Recording

The MaxOne system (MaxWell Biosystems) was used to stimulate and measure neuronal activity in the dissociated cultures. The MaxOne array contains a total of 26,400 microelectrodes arranged in a 220 × 120 grid over a test area of 3.85 mm × 2.1 mm. Of these, 1,024 electrodes can be used simultaneously for measurement. Each electrode measures 9.3 μm × 5.45 μm, with an inter-electrode spacing of 17.5 μm, and the measurement sampling frequency is 20 kHz.

To select appropriate recording and stimulation sites, we first performed a comprehensive scan of the entire culture to identify the regions of spontaneous neuronal activity. Short recording sessions were performed in patches across the array, allowing us to map the neuron-dense regions of significant activity. Based on this initial scan, 1,024 electrodes were selected from the active regions of the array for recording, prioritizing electrodes positioned near clusters of neurons (see Figure 1B for an example of the firing rate).

Following this initial scan-based selection of recording electrodes, a 30-minute continuous recording session of spontaneous activity was performed on the selected 1,024 electrodes. This session provided a detailed assessment of the spike amplitude and firing rate at these sites, allowing for a more refined analysis of neuronal activity.

For the stimulation electrodes, a subset of 6 electrodes was selected from the high-activity recording electrodes, with an additional spacing requirement between each selected stimulation site to prevent cross-activation. Electrical stimulation was delivered using biphasic voltage pulses with a pulse width of 200 μs and a positive leading amplitude of 350 mV.

## 2.3 Experimental Paradigms

To investigate SSA and deviance detection in cultured neurons, we used four main stimulus paradigms: the oddball paradigm, the omission paradigm, the many standards control (MSC) paradigm, and the predictable deviant paradigm. In each paradigm, stimuli were categorized as either standard or deviant. The standard stimulus was the frequently presented stimulus designed to create a familiar, repetitive pattern, while the deviant stimulus was the infrequently presented or omitted stimulus introduced to disrupt the regular sequence and allow assessment of neuronal responses to unexpected events. Each paradigm was conducted using a "flip-flop" design, in which the roles of standard and deviant stimuli were alternated over two runs. This design enabled the direct comparison of neuronal responses to the same stimulus locations in different roles, effectively controlling for location-specific effects.

Across the main paradigms, we defined ten different stimulation locations (labeled Stim A to Stim J). For the oddball and omission paradigms, a subset of three locations selected from these ten was used consistently. The roles of these three selected locations were swapped across runs to create six unique combinations of standard, deviant, or omitted stimuli. In the MSC condition, however, all ten stimulus locations were used, with each presented in a randomized order with a 10% probability. This configuration provided an equal-probability baseline, thus providing a balanced comparison to differentiate deviance detection from simple adaptation effects (see Figure 1C for configurations). In all conditions, stimuli were presented 600 times at 500 ms intervals. In the oddball paradigm, the three selected locations were presented in a randomized 9:1 ratio of standard to deviant stimuli. In the second run, the roles of the standard and deviant locations were reversed, allowing us to examine neuronal adaptation by comparing responses when locations alternated between frequent (standard) and infrequent (deviant) roles. After each run, the cultured neurons rested for 5 min to reduce the effect of the previous stimulus set on the next run.

The omission paradigm also used a 9:1 ratio, using the same three locations as in the oddball paradigm for standard stimulation, but replacing the deviant stimulus with an omitted event in the sequence. This design allowed us to explore neural adaptation to unexpected omissions within a structured stimulus sequence.

In the MSC condition, each of the ten defined stimulus patterns (Stim A-J) was presented with equal probability, each at 10% in a randomized sequence. This setup provided an equal probability baseline, allowing direct comparison with the oddball paradigm to assess adaptation in the absence of deviance-related rarity.

The predictable deviant paradigm was conducted in separate cultures from those used in the other paradigms. In this paradigm, three unique stimulus locations were selected, and sequences alternated between periodic and random conditions with three probability settings (5%, 10%, and 20%) for the deviant stimulus. In the periodic condition, the deviant stimulus appeared at fixed intervals, creating a predictable sequence, whereas in the random condition, the sequence was completely randomized while maintaining the same overall probability. The same three locations were consistently used for both the periodic and random conditions, and no MSC or omission paradigms were used in this setup. This design allowed us to assess how different levels of predictability affect neural adaptation and deviance detection (see Figure 1C for configuration details).

## 2.4 Pharmacology

In a subset of experiments, the NMDA receptor antagonist D-2-amino-5-phosphonovaleric acid (D-AP5; Tocris) was used to block synaptic transmission to examine the effect of synaptic transmission on deviance detection.

The D-AP5 stock was diluted in the solution to the final concentration (50 μM) prior to application. Drugs were applied to cultures and were allowed to equilibrate for at least 20 minutes before the recording began to avoid any putative interference from transient changes in the network activity that might have been induced by culture handling. After the D-AP5 experiment, the entire medium was gently replaced three times to minimize residual D-AP5. Subsequent to the removal process, the culture was allowed to stabilize for at least 20 minutes before initiating the recording.

## 2.5 Data analysis

In addition to the experiment with different inter-stimulus intervals (ISIs), the stimuli were presented 600 times with 500 ms ISIs in all conditions. The measured potential data were passed through a bandpass filter from 300 to 3000 Hz to detect the action potentials. The 1 ms post-stimulus period was excluded from the firing detection range to eliminate the influence of stimulation artifacts. To exclude the effect of preference for different stimuli by cultured neurons, three different stimulation groups were used for each experimental paradigm. Finally, the group with the largest response amplitude to stimulation was selected for analysis. Firing rates of responses to stimuli were calculated from peri-stimulus time histograms (PSTH) with a time bin of 1 ms.

To assess the deviance detection capabilities of the cultured neurons, we analyzed their responses to standard, deviant, and MSC stimuli. The MSC condition presented multiple stimuli with the same overall frequency as the deviant stimulus but at different locations, allowing us to distinguish true deviance detection from the SSA.

We quantified SSA using the SSA index (SI):

$$SI(R) = \frac{R - R_{std}}{R + R_{std}},$$

where $R$ represents the response to either deviant ($R_{dev}$) or MSC ($R_{MSC}$) stimuli, and $R_{std}$ represents the response to standard stimuli. A positive SI indicates that the response to the deviant or MSC stimulus exceeds that response to the standard stimulus. A higher SI for deviant stimuli than MSC stimuli indicates true deviance detection beyond mere adaptation.

To directly compare the SI of the deviant and MSC, we defined the Deviance Detection Index (DDI) as

$$DDI = \frac{R_{MSC} - R_{dev}}{R_{MSC} + R_{dev}}.$$

The DDI intuitively represents the relationship between deviant and MSC responses. If the DDI is positive, it means that there is true deviation detection.

## 3 Results

### 3.1 Cultured Neurons Exhibit Enhanced Late Responses to Deviant Stimuli

To investigate how neural circuits process unexpected inputs, we recorded responses from dissociated rat cortical neurons grown on high-density CMOS arrays under various stimulation paradigms (Figure 1). Comparing firing rates between standard (frequent) and deviant (infrequent) stimuli allowed us to assess MMR in vitro. Raster plots showed that deviant stimuli elicited larger responses than standard stimuli (Figure 2A). In line with this observation, population peri-stimulus time histograms (p-PSTHs)—averaging firing rates across all channels and trials—indicated significantly stronger responses to deviant stimuli (Figure 2B), consistent with an MMR in vivo characterized by enhanced neural activation to unexpected events.

Our analysis revealed two distinct response phases: an early phase (0-10 ms post-stimulus) and a late phase (11-150 ms post-stimulus), each visible as a separate peak in the mean p-PSTH (Figure 2B). This two-phase pattern is consistent with previous studies on cultured networks, in which the early phase primarily reflects direct electrical activation and the late phase corresponds to synaptically mediated network recruitment (Jimbo et al., 2000; Eytan and Marom, 2006; Bakkum et al., 2008; Kermany et al., 2010; Gal et al., 2010; Dranias et al., 2013). Although individual electrode channels varied in their early-phase responses (Supplementary Figure 1A-C), the late-phase responses to deviant versus standard stimuli consistently showed larger amplitudes than the early responses. Because the early responses were likely dominated by direct stimulation rather than synaptic processes, we focused on the late (11-150 ms) responses for further analysis of deviance detection (see Discussion 4.1 for further details).

### 3.2 Cultured Neurons Exhibit Robust Deviance Detection Properties

To distinguish true deviance detection from SSA, we compared late-phase (11-150 ms) responses to deviant stimuli with those in the MSC condition. The MSC condition presents multiple stimuli with equal probability, ensuring that any given stimulus in this condition is rare but not contextually deviant, in the same way as the deviant stimulus in the oddball paradigm. This comparison allowed us to determine whether enhanced deviant responses reflected true deviance detection beyond mere adaptation to stimulus rarity.

The grand average across all cultures revealed that late-phase responses to deviant stimuli were consistently larger than those to MSC stimuli. However, omission as a deviant stimulus failed to elicit a detectable neuronal response in the omission experiment (Figure 3A). To quantify the differences between deviant and MSC responses, we first calculated the SI, which measures how responses differ from standard stimuli. The SI for MSC responses was significantly above 0 in late-phase responses ($p = 1.37 \times 10^{-2}$, Wilcoxon signed-rank test, Figure 3B), indicating basic adaptation to stimulus rarity. Furthermore, deviant stimuli showed larger SI than MSC stimuli in late-phase responses ($p = 1.95 \times 10^{-3}$, Wilcoxon signed-rank test). This difference suggests that deviant stimuli elicited responses beyond those expected from rarity alone. To directly quantify deviance detection, we calculated the DDI, which measures the extent to which larger deviant responses are compared with MSC responses. The consistently positive DDI values (Figure 3C) provide strong evidence that these neural networks perform true deviance detection rather than simply adapting to the stimulus frequency.

The true deviant detection property was observed only in the late-phase but not in the early-phase responses (Supplementary Figure 1D). The SI for early MSC responses was significantly above 0 ($p = 1.95 \times 10^{-3}$, Wilcoxon signed-rank test), indicating the presence of adaptation even in the earliest response window. However, unlike in the late responses where the SI of the deviant was significantly larger than that of the MSC, no such difference was found in the early response window. This finding suggests that while adaptation occurs in early responses, true deviance detection requires later processing phases.

The time windows of deviance detection were different from those of SSA in neural responses (Supplementary Figure 2). At the level between cultures, no significant difference between deviants and other responses was observed in the first 10 ms post-stimulus latency, yet MSC responses were larger than standard responses due to SSA (Wilcoxon signed-rank test with Bonferroni corrections, $p = 5.86 \times 10^{-3}$). Deviant responses then became significantly larger than standard responses at 10 ms or later (Wilcoxon signed-rank test with Bonferroni corrections, $p = 5.86 \times 10^{-3}$, for 10-20 ms), and larger than MSC responses at 20 ms or later ($p = 1.76 \times 10^{-2}$ for 20-30 ms). These results suggest that SSA is observed in the first 10 ms post-stimulus latency, whereas the true deviance detection develops after 20 ms.

Corresponding to these temporal patterns, the peak latency of deviant responses (24.5 ± 12.8 ms) was also later than that of standard (14.1 ± 4.28 ms) and MSC responses (14.0 ± 3.85 ms, Table 1) during late-phase responses. Furthermore, the response duration, defined as the post-stimulus period in which the response amplitude exceeded three standard deviations above the average within the 50 ms preceding stimulus onset, was 161 ± 48.0 ms for deviant stimuli, which was longer than that of MSC (147 ± 55.7 ms) and shorter than that of standard stimuli (174 ± 46.5 ms).

The standard responses decreased rapidly at first, then gradually with stimulus repetition due to adaptation, while the deviant and MSC responses remained stable throughout the experiments (Figure 4A). Statistically, the first standard stimulus elicited a significantly larger response than the last standard stimulus, and the average of the standard responses in the first half was significantly larger than that in the second half, while the deviant and MSC responses showed no significant changes (Figure 4B). Throughout the experiments, deviant responses were larger than MSC responses, suggesting that deviance detection depends more on contextual patterns than stimulus frequency.

### 3.3 NMDA Receptor-Dependent Synaptic Transmission is Required for Deviance Detection

To determine whether deviance detection depends on NMDA receptor-mediated synaptic transmission, we administered the NMDA receptor antagonist, D-AP5, and compared neuronal responses in three conditions: before D-AP5 application (baseline, Figure 5A), during D-AP5 treatment (Figure 5B), and after D-AP5 washout (Figure 5C). Following D-AP5 application, the overall response duration became shorter (standard: 47.0±32.2 ms; deviant: 37.7±19.6 ms; MSC: 23.0±7.48 ms), with a marked decrease in the late-phase responses (11-150 ms), which became indistinguishable between conditions (standard: 0.0116±0.0116 ms; deviant: 0.0164±0.0068 ms; MSC: 0.0130±0.0090 ms). Notably, early responses (0-10 ms) showed a slight increase in average amplitude during D-AP5 treatment. After the D-AP5 washout, the responses recovered to pre-D-AP5 baseline levels (Figure 5C, Table 2). No omission responses were observed throughout the experiments.

The selective suppression of late responses during NMDA receptor blockade, while early responses persisted, provides strong evidence that the late-phase responses are critically dependent on synaptic transmission associated with the later, longer-lasting NMDA receptor. The responses observed following DAP-5 administration, which did not exhibit deviance detection, are likely mediated primarily by the early, rapid AMPA receptor component (Jimbo et al., 2000; Kielland & Heggelund, 2001). These results suggest that deviance detection, which manifests primarily in the late phase, relies on NMDA receptor-mediated synaptic signaling rather than the direct electrical activation of neurons.

### 3.4 Spatial Distribution of Neural Responses and Spatiotemporal Response Propagation

We examined whether the early- and late-phase responses depended on the distance from the stimulation sites. In the early responses, both standard and deviant stimuli evoked larger responses near the stimulation sites (i.e., within a 15-electrode radius; Figure 6A) than those in distant regions (Figure 6B, left; Wilcoxon signed-rank test with Bonferroni corrections, $p=1.17\times10^{-2}$ for standard vs. other areas in response to standard stimulus; $p=1.17\times10^{-2}$ for deviant vs. other areas in response to deviant stimuli). In contrast, no significant differences were found in the spatial distribution of the late-phase responses, suggesting that early local activation of neurons near the stimulation sites is followed by progressive recruitment of distant neurons.

To further characterize how the neural activation spread as a function of distance from the stimulation site (Figure 7A), we categorized the recording sites into 4 groups according to the distance from the stimulation site (Figure 7B; 15-, 30-, 45-, and 60-electrode radii) and examined p-PSTH for each group in response to standard and deviant stimuli (Figure 7C). The early responses were largest near the stimulation site (within the 15-electrode radius) and decayed with the distance from the stimulation site in response to both standard and deviant stimuli (Figure 7D; Wilcoxon signed-rank test, $p=3.13\times10^{-2}$ for 15- vs. 60-electrode radius in standard; $p=3.13\times10^{-2}$ in deviant). On the other hand, no significant distance-dependent difference was observed in the late-phase responses (Figure 7E). These results suggest that rapid local activation was followed by network-wide activation.

### 3.5 Unpredictable Temporal Patterns Enhance Responses to Deviants and Standards

To investigate how temporal predictability affects neural activity, we compared stimulus sequences in which deviants appeared either randomly or periodically with the same overall probability (10%) (Figure 8A). Across more than 8,000 recorded sites, all responses within each culture were normalized to z-scores to account for baseline firing differences.

Focusing on the late-phase window (11–150 ms), we found that both standard and deviant stimuli evoked higher firing rates when presented in the random sequence compared to the periodic sequence. Figure 8B displays representative plots of mean neuronal responses from individual cultures, visually demonstrating stronger responses to deviants and standards in the random condition. A paired Wilcoxon signed-rank test for quantitative analysis of overall individual sites was conducted for further comparison. This test revealed a significant difference for standard responses (n=7,995, $p=7.51\times10^{-151}$), with the random condition eliciting stronger responses overall (Figure 8C, left; mean z-score under random vs. periodic: $0.050 \pm 0.005$ vs. $0.000 \pm 0.005$). Based on the tests for individual Wilcoxon signed-rank tests performed for each site, 19.5% of sites exhibited significantly stronger standard responses to random sequences, while only 5.3% preferred periodic sequences (Figure 8D, top). Deviant responses showed a similar late-phase trend. A significant difference overall (n=7,820, $p=2.40\times10^{-14}$), with a trend towards stronger deviant responses for random deviants (Figure 8C, right; mean z-score under random vs. periodic: $0.118 \pm 0.005$ vs. $0.076 \pm 0.005$). The individual tests for each site showed that 13.4% of sites preferred random deviant stimulus, while 6.5% preferred periodic deviant stimulus (Figure 8D, bottom).

Collectively, these results show that random, unpredictable stimulus orders robustly enhance late neural responses for both standards and deviants, echoing the sensitivity to a result that parallels the sensitivity to sequential contexts observed in vivo (Yaron et al., 2012). Because late responses are predominantly driven by synaptic transmission, this enhancement supports the notion that network-level processing is particularly sensitive to irregular sequences, while more predictable periodic inputs elicit comparatively weaker activation.

## 3.6 Spatial and Temporal Parameters Shape Network Adaptation and Deviance Detection

To characterize how neural networks adapt to and process stimulus patterns, we investigated the spatial configuration, recovery dynamics, and temporal parameters of deviance detection.

We first examined how the spatial separation between the standard and deviant stimulus electrodes affected neural responses (Figure 9A and Supplementary Figure 3A). During early responses, no significant firing rate difference was observed between standard-deviant pairs regardless of separation distance (Wilcoxon signed-rank test; Supplementary Figure 3B). During late responses, the responses to deviant stimuli were not significantly different from those to standard stimuli when the spatial separation was small (18-40 μm; Wilcoxon signed-rank test, p=0.12), but became significantly larger than those to standard when the spatial separation was medium (41-68 μm; $p=4.09\times10^{-2}$) and large (69-175 μm; $p=9.46\times10^{-3}$) (Figure 9B). These results suggest that the spatial separation between stimulus locations enhances deviance detection, possibly because the difference between the deviant and standard stimuli becomes more distinct with increasing separation.

Neural networks also require specific recovery periods to restore their response amplitudes after adaptation. After providing 20 identical stimuli at 500-ms ISI, we characterized neural responses to the 21st stimulus, which was presented with a given recovery time ranging from 0.5 s to 10.5 s. Our data showed that the neural responses increased with a recovery time of at least up to 10 s (Figure 10A). When a different stimulus was presented within this recovery time, the recovery was less effective than when no stimulus was presented.

The influence of repeated standard stimuli on deviance detection was then systematically examined as a function of the number of standard stimuli presented before each deviant stimulus. The deviant responses increased with the number of preceding standard stimuli, reaching a maximum amplitude after 15 consecutive standard presentations (Figure 10B and C). This result suggests that neural networks progressively strengthen their response to deviations with increasing exposure to standard patterns and that this memory capacity is at least 15 stimuli or 7.5 s at 500 ms ISI.

Investigations across multiple timescales revealed remarkable temporal flexibility in deviance detection. Using three complementary paradigms (oddball, omission, and MSCs), we tested the effects of ISIs on neural responses. The average p-PSTHs showed consistent deviant enhancement at ISIs of 100 ms (Figure 11A), 500 ms (Figure 11B), and 1000 ms (Figure 11C). Quantification of the SI confirmed that the neural networks maintained their ability to detect deviant stimuli at 500 ms (Wilcoxon signed-rank test, $p=7.81\times10^{-3}$) and 1000 ms ($p=7.81\times10^{-3}$) (Figure 11D). This preservation of deviance detection at longer intervals indicates that these neuronal cultures can maintain stimulus-specific information for at least 1 second. The optimal ISI for eliciting deviant responses was between 100 ms and 1 s, consistent with a human MMN study (Javitt et al., 1998). ISI affected the response amplitude, yet enhanced responses to deviant stimuli remained consistent across all intervals, suggesting that deviance detection operates independently of presentation rate over at least two orders of magnitude on a temporal scale. Notably, we found no responses to missing stimuli at any tested interval, which differs from observations in human studies where omission responses have been reported.

## 4 Discussion

Our study shows that deviance detection capability and regularity sensitivity emerge spontaneously in simplified neural architectures, without requiring complex brain organization. Using high-resolution CMOS arrays to probe dissociated cortical neurons, we demonstrated that these networks exhibited robust MMRs, particularly in the late phase (11-150 ms post-stimulus), indicating true deviance detection beyond mere adaptation. These MMRs are critically dependent on NMDA receptor-dependent synaptic transmission, and result from the interplay between local response suppression and global network recovery. These findings challenge the traditional view that complex temporal pattern processing requires elaborate neural hierarchies, suggesting instead that sophisticated regularity detection may be an inherent property arising from fundamental principles of neural organization.

### 4.1 Deviant detection in the late-phase response

Numerous studies of dissociated cortical networks indicate that early responses (within approximately 0–20 ms of stimulation) predominantly reflect the direct electrical activation of nearby somas and axons, bypassing critical synaptic and recurrent network processes (Jimbo et al., 2000; Eytan et al., 2003; Bakkum et al., 2008; Gal et al., 2010; Kermany et al., 2010; Dranias et al., 2013). For example, Gal et al. (2010) observed highly precise, short-latency firing immediately after low-frequency stimulation, due to direct depolarization. Similarly, Dranias et al. (2013) reported brief initial spikes (3–20 ms post-stimulus) that did not rely on the reverberatory activity. Such short-latency discharges provide limited insight into deviance detection or other context-sensitive phenomena, as they occur without substantial synaptic integration. In contrast, late-phase responses (lasting tens to hundreds of milliseconds) involve synaptic transmission, recurrent excitation, and adaptive network dynamics (Eytan & Marom, 2006; Bakkum et al., 2008; Gal et al., 2010; Dranias et al., 2013). Bakkum et al. (2008) showed that changes in conduction delay and amplitude over longer timescales require synaptic activity, reflecting a network-level process rather than direct electrode effects. In this 11–150 ms window, neural ensembles exhibit synchronized bursts or wavefronts driven by local excitatory-inhibitory balance and plasticity, which are central to the detection of unexpected inputs (Kermany et al., 2010; Gal et al., 2010). Crucially, deviance detection also emerges beyond the earliest afferent latencies in intact animal preparations (Ulanovsky et al., 2003; Ahmed et al., 2011; Taaseh et al., 2011; Chen et al., 2015; Shiramatsu et al., 2013), supporting the idea that recurrent network processes underlie responses to unexpected stimuli. Thus, focusing on the 11–150 ms window in our cultured neurons captures this synaptic and network-dependent activity, allowing us to distinguish true deviance detection from simple adaptation to repetitive stimulation. Recognizing that these late-phase dynamics depend on synaptic interactions naturally leads to questions about the contributions of NMDA receptors to MMRs.

## 4.2 Deviant detection emerging from synaptic integration

Our pharmacological investigations revealed that NMDA receptor-mediated synaptic transmission was critical for the generation of late MMRs. The selective abolition of late responses while preserving early ones under NMDA receptor blockade demonstrates distinct mechanisms for different processing phases (Jimbo et al., 2000). This aligns with the evidence from recent studies showing reduced mismatch negativity under NMDA receptor antagonism (Shiramatsu et al., 2013; Harms et al., 2018). The dose-dependent nature of NMDA receptor modulation suggests that synaptic plasticity plays a nuanced role in the processing of prediction errors. Tikhonravov et al. (2010) reported similar effects, with low doses enhancing MMRs and higher doses suppressing them. These findings extend recent work demonstrating that NMDA receptor activity is critical for stabilizing memory traces and improving predictions in cultured networks (Lamberti et al., 2024). The interplay between synaptic plasticity and local network fatigue is particularly relevant. Threshold adaptation and synaptic depression—mechanisms highlighted in computational models (Mill et al., 2011; May et al., 2015)—may explain the distinct temporal dynamics observed in our experiments. These mechanisms allow cultured networks to balance local response suppression with global recovery, enabling sophisticated pattern detection even in simplified systems.

The spatial organization of the responses revealed distinct roles for local and global network dynamics in deviance detection. Neural responses that were initially localized near stimulation sites during early phases gradually spread to engage the entire network during late phases. This spatial-temporal dynamic provides insight into how neural circuits integrate local and global information to generate predictive responses (Wagenaar et al., 2006a; Dockendorf et al., 2009). Deviant detection became more pronounced with increasing distance between standard and deviant stimulus locations, consistent with network-wide propagation of deviance-related signals. These findings align with computational models suggesting that distance-dependent network interactions facilitate hierarchical processing (Barascud et al., 2016; Kern & Chao, 2023).

Recent theoretical studies have explored various mechanisms for deviance detection in neural circuits. Our findings, which demonstrate the importance of both synaptic and non-synaptic plasticity complement computational studies that have focused primarily on synaptic mechanisms (Mill et al., 2011; May et al., 2015; Yarden & Nelken, 2017). The interaction between threshold adaptation and synaptic depression in our study is consistent with recent modeling work showing how these mechanisms can work synergistically through complementary effects on local and global network fatigue (Kern & Chao, 2023). In addition, our results support the emerging frameworks of predictive processing in simplified neural circuits and demonstrate basic computational building blocks that could support prediction, which is consistent with recent demonstrations that cultured networks can perform variational Bayesian inference (Isomura et al., 2023) and build predictive models through targeted synaptic modifications (Lamberti et al., 2023).

### 4.3 Deviant detection properties in the dissociated culture of neurons

We showed that the MMRs were generated at inter-stimulus intervals of up to 1 s, demonstrating robust short-term memory capabilities in these simplified networks. This temporal integration window exceeds previous estimates, suggesting that even basic neural circuits can retain information over behaviorally relevant timescales (Dranias et al., 2013; Ju et al., 2015). Interestingly, this timescale is comparable to the optimal inter-stimulus interval to evoke MMN in humans (Javitt, et al., 1998) and is likely to emerge from the time constant of NMDA receptor-mediated synaptic currents (Lester et al., 1990). In addition, the spatial analysis revealed that deviance detection becomes more pronounced with increasing distance between standard and deviant stimulation sites, suggesting that temporal integration involves the progressive recruitment of broader network regions.

These observations are consistent with findings in intact systems, where MMRs are modulated by temporal predictability and inter-stimulus intervals (Näätänen et al., 2012; Recasens et al., 2014). However, in contrast to the results from human EEG experiments (Winkler and Näätänen, 1993), we did not find the omission responses, which may reflect the limitations of dissociated cultures in encoding prediction errors related to absences rather than presences. The generation of omission responses may involve more complicated predictive mechanisms than deviance detection, potentially involving higher-order cognitive processes such as attention. These processes may require a more complex network architecture or a larger neural network for effective manifestation. Furthermore, some studies have suggested that neural circuits can predict omissions based on their internal state; however, an external trigger is required to elicit a detectable omission response, a phenomenon known as state-dependent computation (Buonomano et al., 2009). Consequently, omission responses in vivo can be initiated by multiple pathways within the auditory system, including non-characteristic frequency channels that have been implicated in processing temporal expectations and state-dependent predictive responses.

Our observation that random sequences elicit stronger responses than periodic ones for both deviant and standard stimuli reveals a fundamental property of neural computation: the enhanced processing of unpredictable information. This preference for randomness highlights a core computational strategy in neural circuits to prioritize novel and unpredictable inputs, aligning with studies on the auditory cortex showing enhanced sensitivity to irregularities (Yaron et al., 2012; Mehra et al., 2022). The enhancement of late-phase responses to random sequences suggests that this preference operates at the network level rather than through local circuit mechanisms. Similar preferences for random over predictable patterns have been reported in both in vivo auditory systems and in vitro cultured networks, highlighting the generality of this phenomenon (Southwell et al., 2017; Barascud et al., 2016).

### 5 Conclusion

The present work demonstrates that fundamental mechanisms for deviance detection and temporal pattern processing emerge spontaneously in the dissociated cultures of neurons, one of the most primitive neural architectures. These capabilities appear to be inherent properties of neural networks rather than requiring a specific brain organization and may serve as building blocks for predictive processing. These findings advance our understanding of neural computation and provide new directions for developing more brain-like artificial intelligence systems that incorporate the principles of biological adaptation and pattern recognition.

**Figure legends**

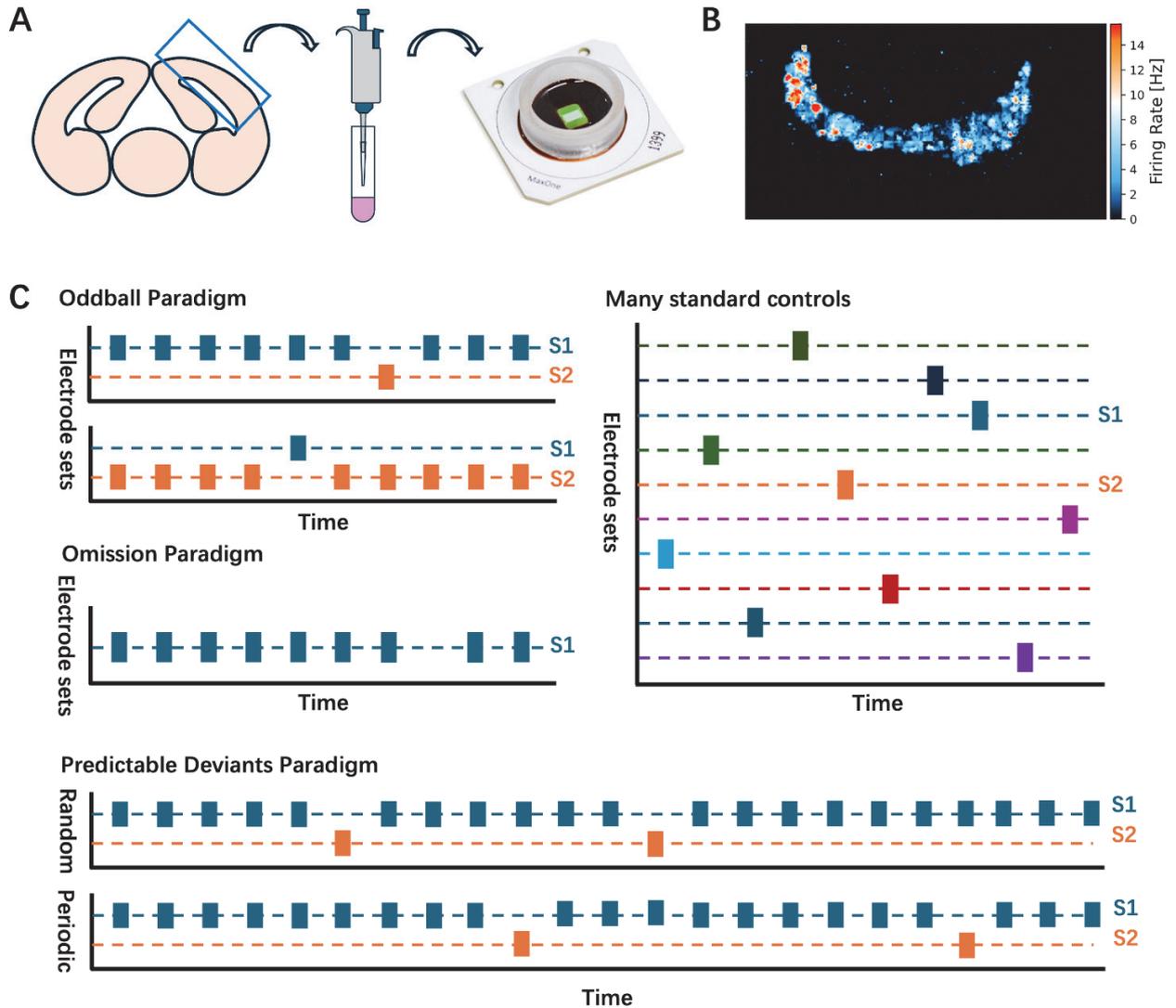

**Figure 1.** Experimental procedure. (A) Material preparation. A schematic diagram illustrates the locations of the cortical neurons utilized in the experiment. Following dissociation, the neurons were plated on the CMOS arrays. (B) Representative spatial map of firing rate in a neuronal culture recorded by CMOS array. (C) Experimental paradigms. Oddball, Omission, MSC, and Predictable deviant paradigms were used in the main experiments. The stimulation electrodes utilized in the oddball and omission experiments were uniform and used in the MSC experiments.

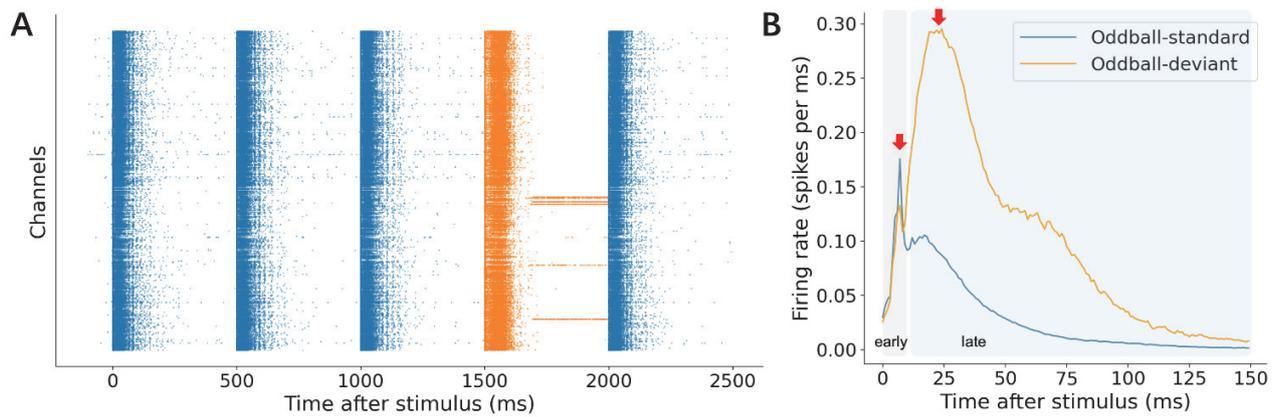

**Figure 2.** Representative mismatch responses (MMRs). (A) A representative raster clip from the oddball experiment. Partial fragments of responses to the deviant stimulus (orange) and responses to the standard stimulus (blue) were selected. Standard and deviant responses represent consecutive stimulus sequences composed of distinct stimuli. (B) Representative p-PSTH. The test stimulus used for standards and deviants were identical to directly compare p-PSTH across conditions. The response was divided into early and late phases each with a different response peak: an early phase, 0-10 ms; and a late phase, 11-150 ms.

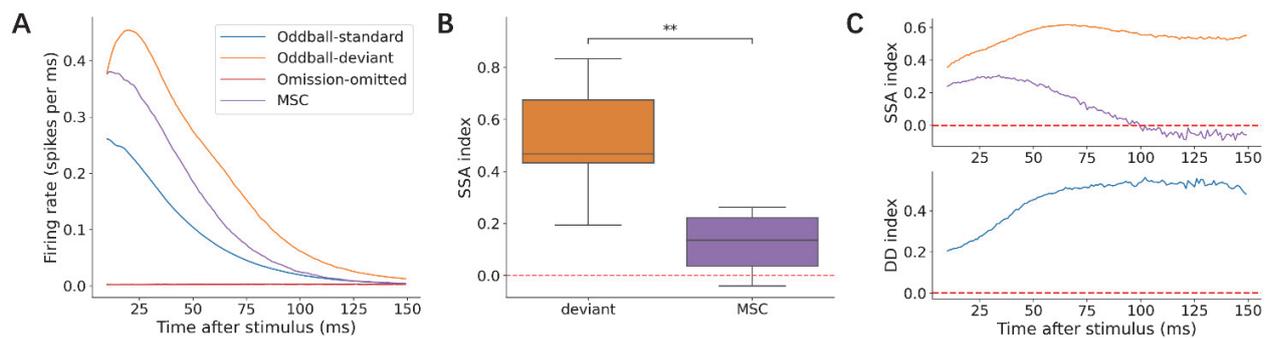

**Figure 3.** Characterization of the late-phase responses. (A) Grand average p-PSTHs in the 11-150 ms post-stimulus latency for oddball, omission, and MSC conditions. (B) SI of the deviant and MSC. In the boxplot here and hereafter, the box shows the median and the upper/lower quartiles, and the whiskers show the minimum/maximum data points except for outliers. Outliers are defined as 1.5 times greater than the upper/lower quartiles. (C) SSA index and DDI as functions of post-stimulus latency are shown.

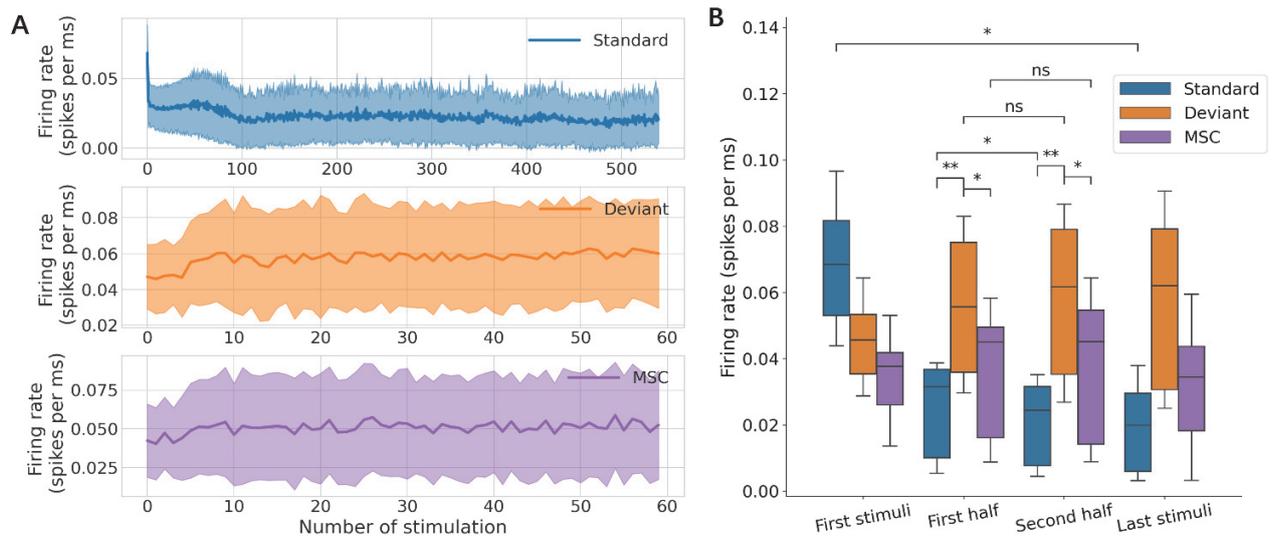

**Figure 4.** Effects of repetition. (A) Response amplitudes were tracked across the complete stimulation sequence of 5 min, in which 540 standard and 60 deviant stimuli were provided. (B) Response amplitudes for each stimulation condition. Neural responses were characterized in the first stimulus, first half, second half, and final stimulus.

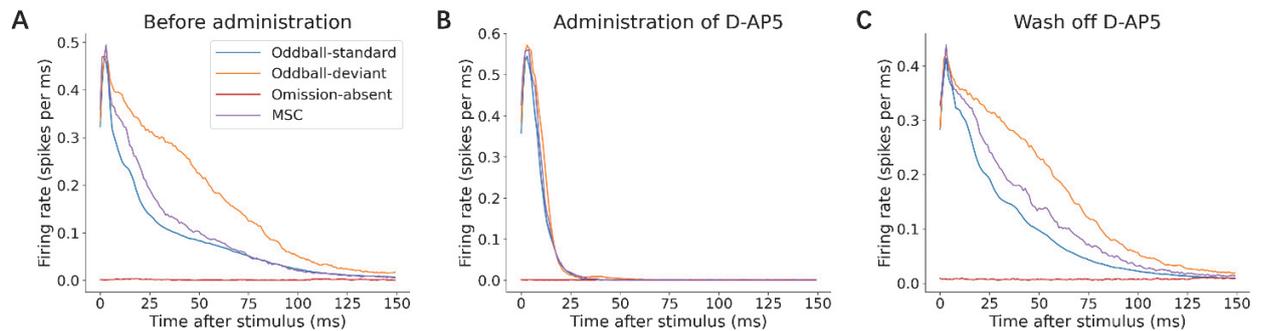

**Figure 5.** Effects of NMDAR. P-PSTHs (A) before and (B) after D-AP5 administration and (C) after D-AP washout. The late responses were eliminated by D-AP5, while recovered after washing off the D-AP5.

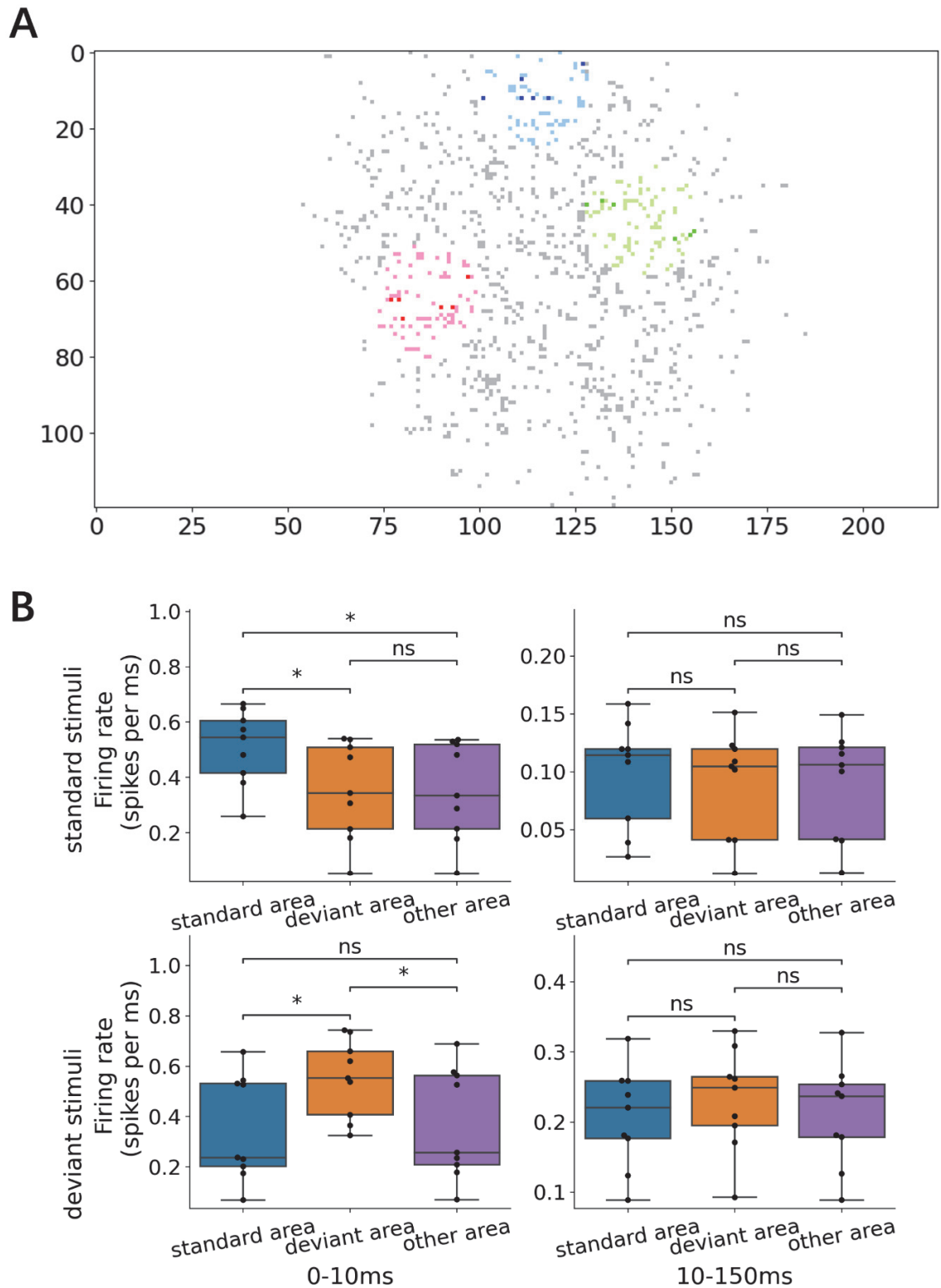

**Figure 6.** Local early response and network-wide late response. (A) Representative map of the recording electrodes. Dark dots denote the stimulating electrode while the surrounding light dots
16

mark locations 15 electrode lengths away, defined as the area proximal to the stimulation electrode (standard, deviant, and other areas in inset B). (B) Responses depending on the proximity of the stimulation electrode: left, early responses; right, late responses; upper standard stimuli; and lower, deviant stimuli. For both early and late responses, the early response was larger in the proximity of the stimulation electrode than in other distant areas, i.e., local activation, whereas the late responses were not dependent on the proximity, i.e., network-wide activation.

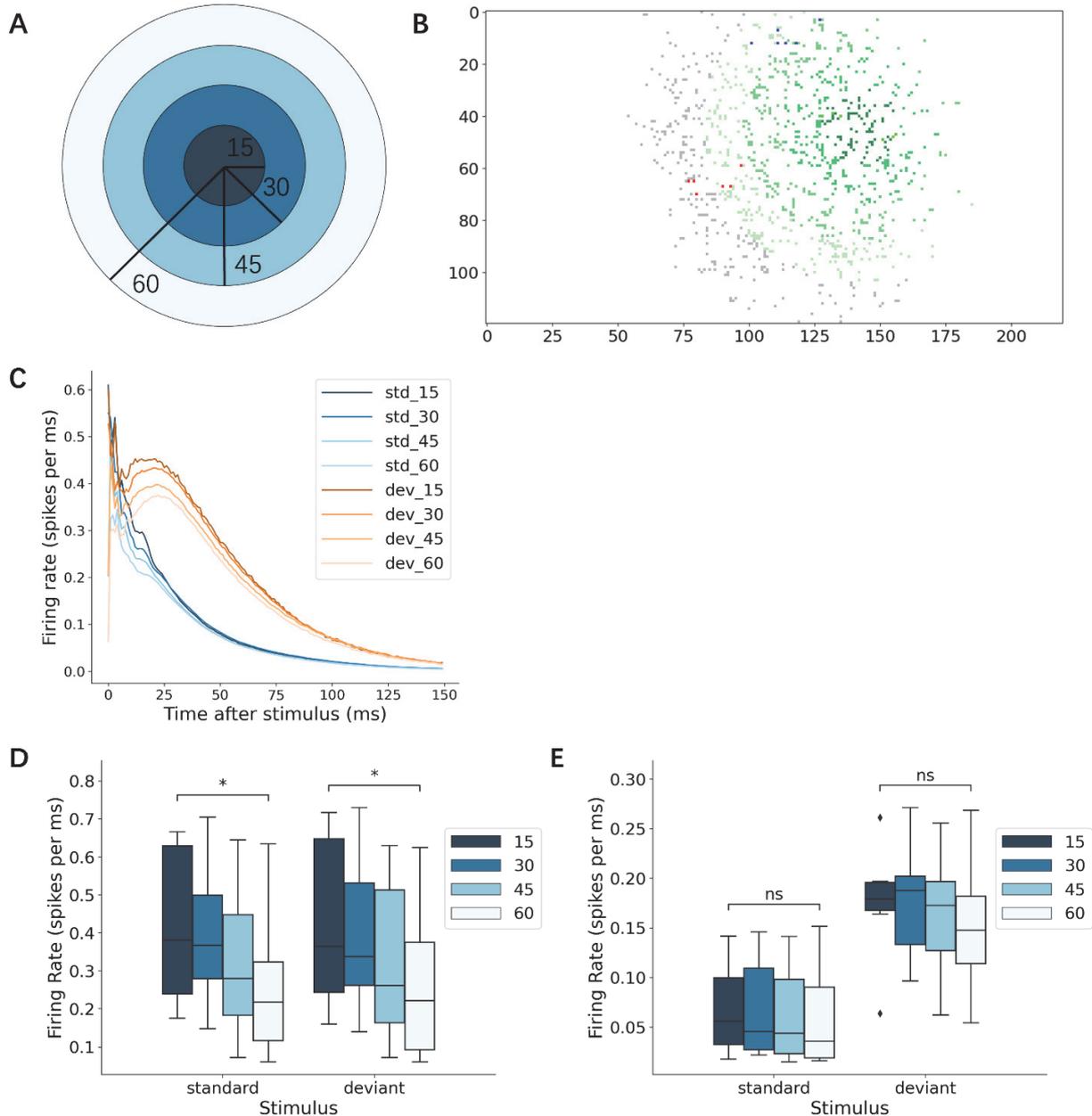

**Figure 7.** Spatial spread of neural responses. (A) Schematic diagram depicting the electrode distance from the stimulus center. In the following analyses, recording electrodes were categorized into 4 groups according to the distance from the stimulation electrode: 15, 30, 45, and 60 electrodes. (B) Representative categorization of recording electrodes. Colors indicate electrode groups according to the distance from the stimulus center. Red, blue, and green electrodes correspond to three different stimulation electrode groups, with green electrodes serving as the stimulation electrodes in this example. (C) Responses to standard and deviant stimuli at varying distances from the stimulation electrodes. The outward spread of neuronal responses was shown. (D) Early responses to standard and deviant stimuli at various distances. Early responses decayed with the distance: the proximal electrodes (within 15 electrode distance) exhibited larger responses than the distant electrodes (at 60 electrode distance). (E) Late responses to the standard and deviant stimuli at different distances. No distance dependence was observed.

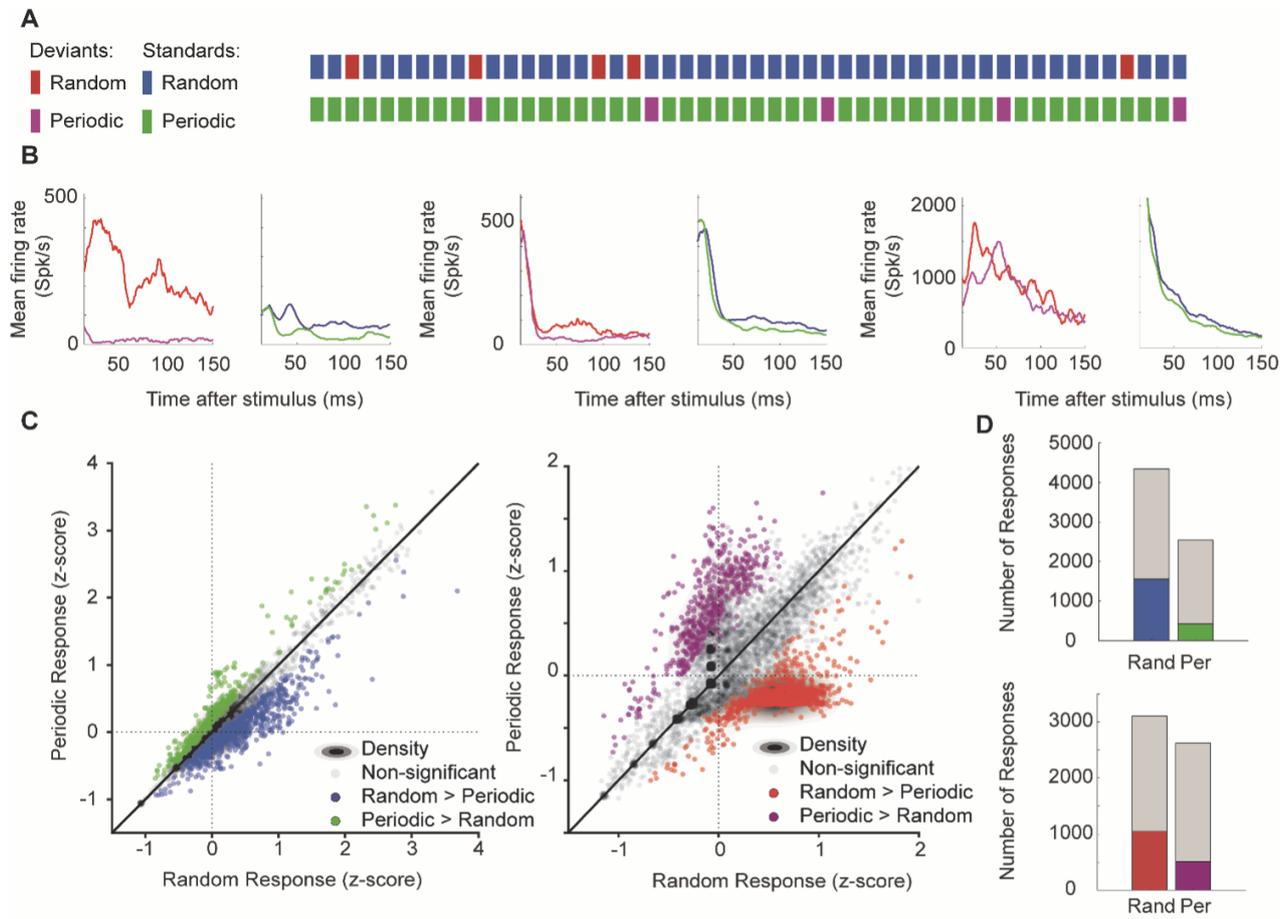

**Figure 8.** Effects of stimulus predictability. Unpredictable stimuli enhanced the late-phase responses. (A) Schematic representation of stimulus sequences showing random and periodic patterns for both deviants (red/purple) and standards (blue/green). (B) Representative p-PSTHs showing mean firing rates in response to random versus periodic sequences for both deviant and standard stimuli across different recording sites. (C) Scatter plots comparing z-scored responses to random versus periodic sequences: left, responses to standard stimuli; right, responses to deviant stimuli. Gray dots indicate non-significant differences and colored dots show significant response preferences (green, periodic > random; blue/red, random > periodic). Density contours indicate the response distributions. (D) Response preferences. The number of recording sites with significant preferences for random (blue/red) versus periodic (green/purple) sequences are shown for standards (top) and deviants (bottom). Gray portions indicate non-significant responses.

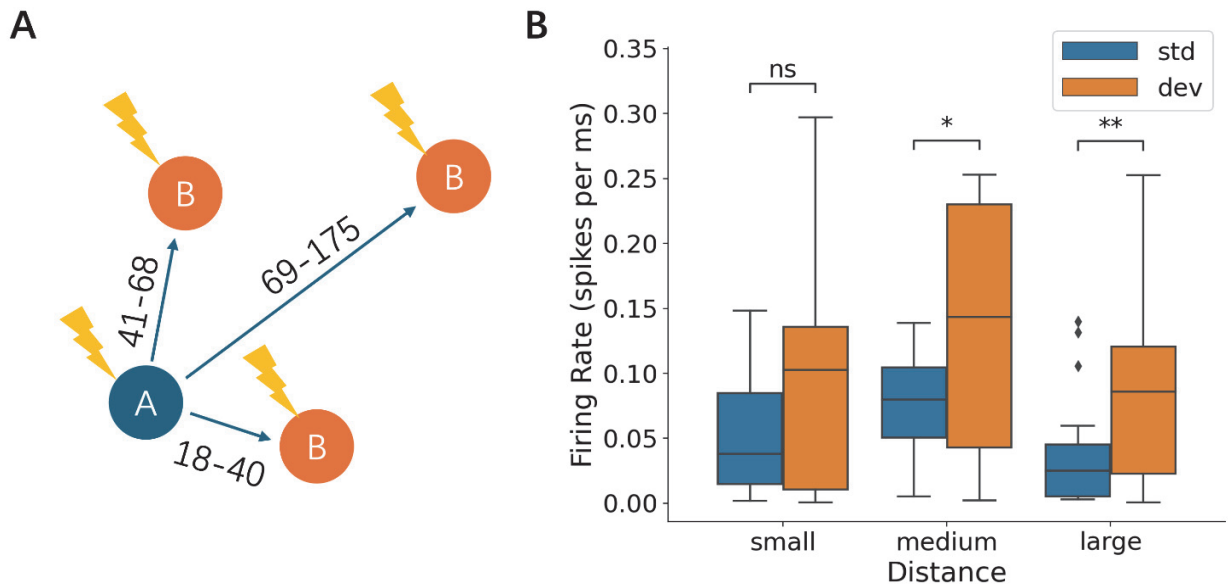

**Figure 9.** Effects of spatial separation between stimulation electrodes. (A) Schematic diagram of the spatial separation between the standard and deviant stimulation electrodes. Separation was categorized into small (18-40 μm), medium (41-68 μm) and large (69-175 μm) distances. (B) Late-phase responses depended on the spatial separation of the electrodes. The difference between deviant and standard responses was not significant when the distance was small, but became significant when the distance was medium or large.

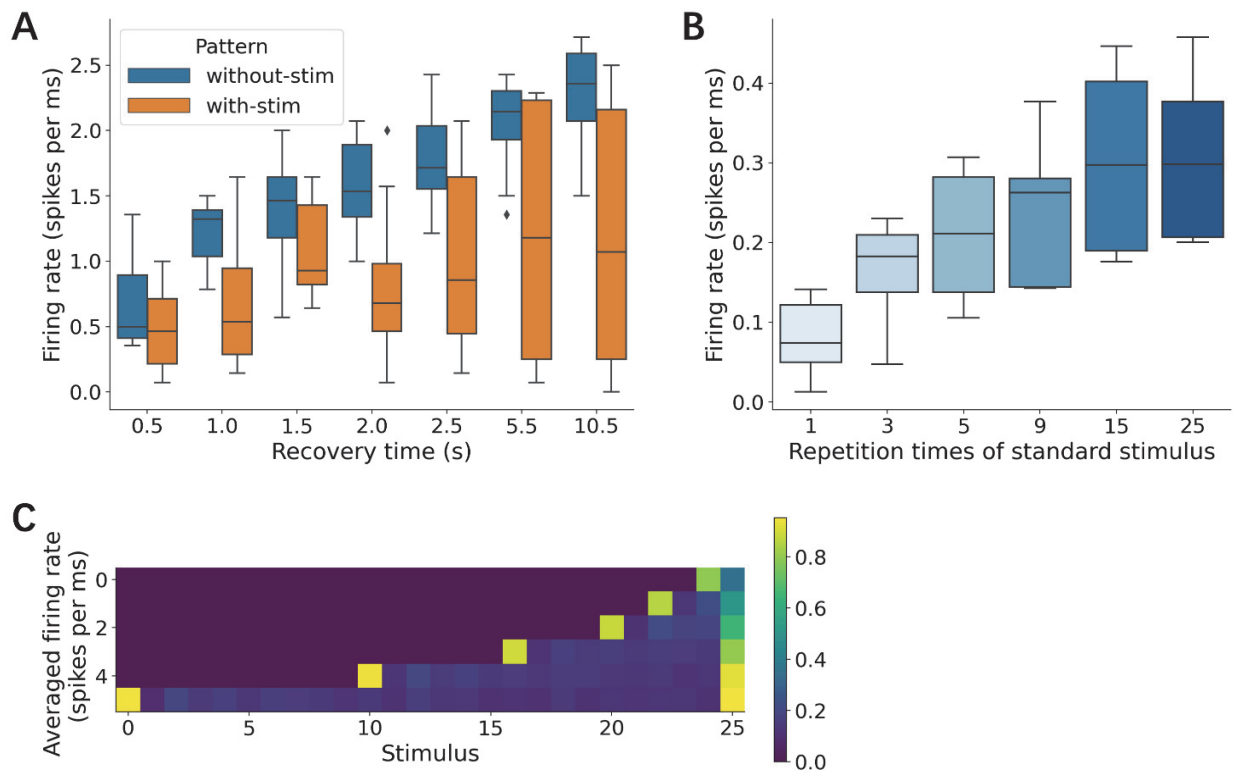

**Figure 10.** Effects of resting time and deviant stimulus frequency. (A) Resting time. Neural responses were characterized immediately after the rest, followed by 20 repetitive stimuli. During the rest, two conditions were set, complete rest (no stimulation) and exposure to different stimuli. (B) Repetition times of standard stimuli. Deviant response increased with the repetition times. (C) Averaged responses to stimulus sequence. Each row indicates different conditions of repetition times of standard stimulus. The most left collum shows the neural responses to a deviant.

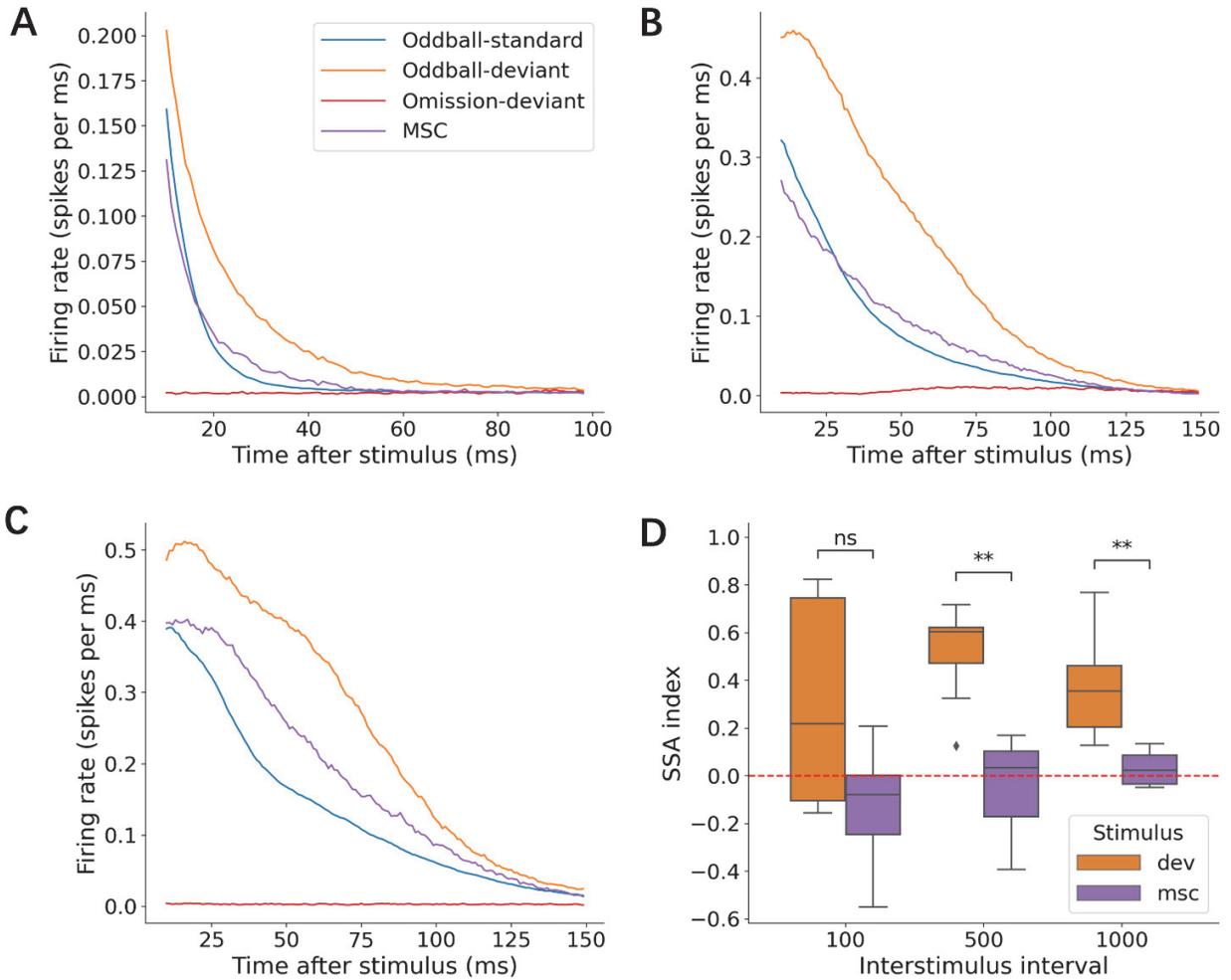

**Figure 11.** Effects of ISI. Average p-PSTHs are shown when ISI was (A) 100ms, (B) 500ms, and (C) 1000ms. (D) SSA index for each ISI.

**Tables**

**Table 1.** Comparison of peak latency and response duration across standard (STD), deviant (DEV), and MSC stimuli in late-phase responses.

|               | STD  | DEV  | MSC  |
|---------------|------|------|------|
| **Peak-Mean (ms)** | 14.1 | 24.5 | 14.0 |
| **Peak-SD (ms)**   | 4.28 | 12.8 | 3.85 |

| | | |
|---|---|---|
| **Duration-Mean (ms)** | 174 | 161 | 147 |
| **Duration-SD (ms)** | 46.5 | 48.0 | 55.7 |

**Table 2.** Peak latency, response duration, and average amplitude in the late-phase responses. Mean ± SD are shown before/after D-AP5 administration and after D-AP5 washing off.

| | | Peak latency (ms) | Duration (ms) | Average amplitude (spikes per ms) |
|---|---|---|---|---|
| **Before (Baseline)** | STD | 13.3±3.3 | 204±54.9 | 0.0638±0.0388 |
| | DEV | 11.0±0 | 171±36.3 | 0.144±0.0368 |
| | MSC | 11.3±0.47 | 133±57.3 | 0.0775±0.0463 |
| **D-AP5** | STD | 11.0±0 | 47.0±32.2 | 0.0116±0.0116 |
| | DEV | 11.0±0 | 37.7±19.6 | 0.0164±0.0068 |
| | MSC | 11.0±0 | 23.0±7.48 | 0.0130±0.0090 |
| **Wash** | STD | 14.7±3.3 | 124±44.1 | 0.0749±0.0651 |
| | DEV | 22.0±14.9 | 171±42.1 | 0.145±0.0787 |
| | MSC | 13.7±3.77 | 124.0±37.7 | 0.102±0.0679 |

### Ethics statement

All animal experimental protocols were approved by the Graduate School of Information Science and Technology's Ethics Committee at the University of Tokyo (JA21-9). All experimental procedures were carried out in accordance with these approved guidelines.

### Conflict of Interest

The authors declare that the research was conducted in the absence of any commercial or financial relationships that could be construed as potential conflicts of interest.

### Author Contributions

ZZ: Conceptualization, Investigation, Methodology, Data curation, Formal analysis, Visualization, Writing – original draft. AY: Conceptualization, Investigation, Methodology, Data curation, Formal analysis, Visualization, Supervision, Validation, Writing – original draft, Writing – review & editing. DA: Investigation, Methodology, Resources, Supervision, Validation, Funding acquisition, Writing – review & editing. TS: Methodology, Funding acquisition, Writing – review & editing. ZC: Validation, Writing – review & editing. HT: Conceptualization, Funding acquisition, Project administration, Resources, Supervision, Validation, Writing – review & editing.

**Funding**

This work is partly supported by JSPS KAKENHI (23H03465, 24H01544, 24K20854),  AMED (24wm0625401h0001), JST (JPMJPR22S8), the Asahi Glass Foundation, and the Secom Science and Technology Foundation.

**Acknowledgments**

The authors acknowledge the use of ChatGPT (OpenAI, version 4o) for assistance in language editing.

**Data Availability Statement**

The raw data supporting the conclusions of this article will be made available by the authors, without undue reservation.

Supplementary Information for

# Deviance Detection and Regularity Sensitivity in Dissociated Neuronal Cultures

Zhuo Zhang, Amit Yaron, Dai Akita, Tomoyo Isoguchi Shiramatsu, Zenas C. Chao, Hirokazu Takahashi

Correspondence to: Hirokazu Takahashi

Corresponding author. Email: takahashi@i.u-tokyo.ac.jp

**This PDF file includes:**

Supplementary Figure 1 to 3

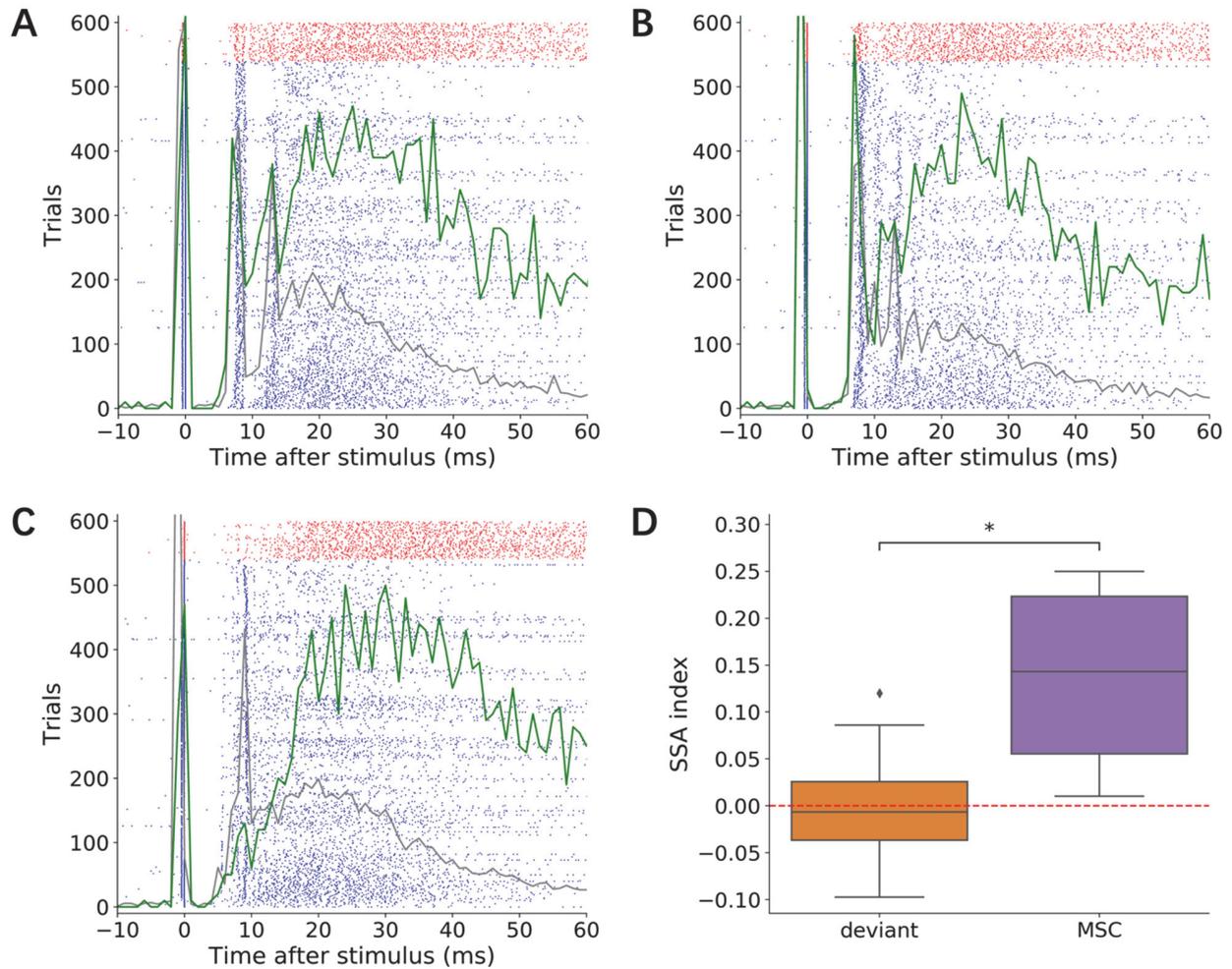

**Supplementary Figure 1.** Early responses. (A) Example of neural responses where the early response to the standard stimulus is equal to the deviant response. (B) Example showing a smaller early response to the standard stimulus compared to the deviant response. (C) Example displaying a greater early response to the standard stimulus than that to the deviant response. The raster plots depict neuronal responses to standard (blue dots) and deviant (red dots) stimuli recorded from a single electrode channel. The lines represent the averaged responses for standard (gray) and deviant (green) stimuli. (D) SSA index of early-phase responses (0–10 ms).

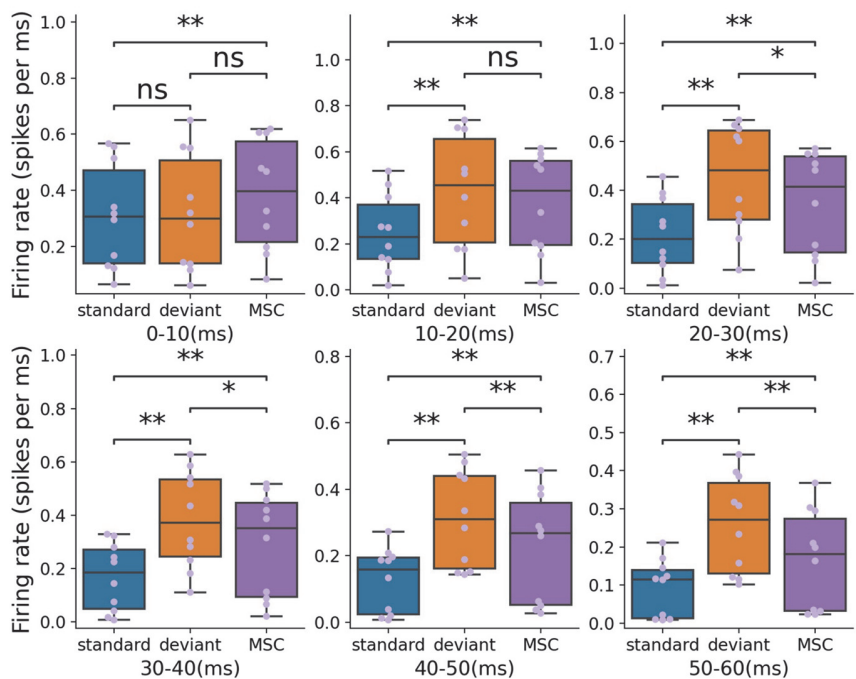

**Supplementary Figure 2.** Deviance detection in late responses. The box plots showed the responses to stimulus in different time windows at 0-10 ms, 15-30 ms, 30-40ms, 40-50 ms, and 50-60 ms. SSA was observed at the first 10ms, while deviant detection after 30ms.

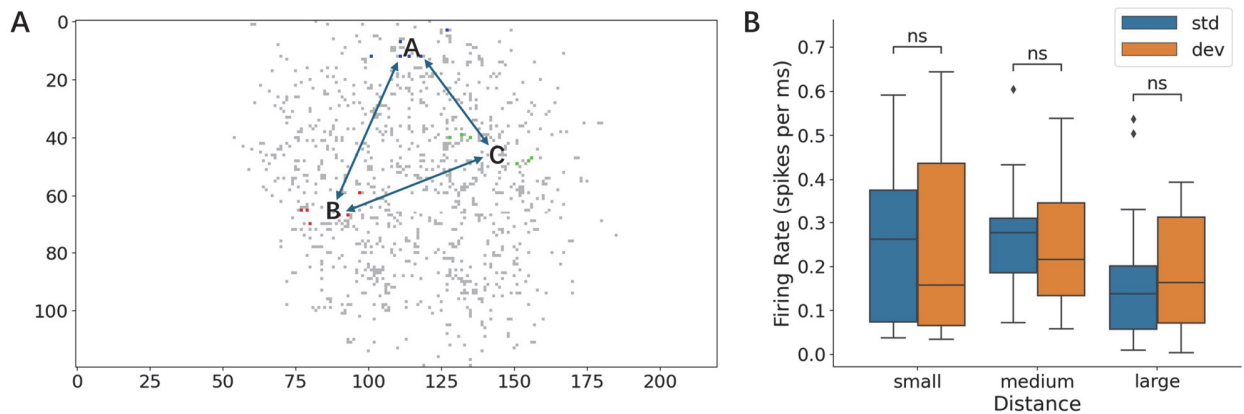

**Supplementary Figure 3.** Distance between stimulation electrodes did not affect early-phase responses. (A) Example of the distance between standard and deviant stimulation electrodes. (B) Early responses to standard and deviant stimuli as a function of the distance between standard and deviant stimulation electrodes.